\documentclass{article}
\usepackage{protools-arxiv}
\def\QED{\hfill\ensuremath{\blacksquare}}

\begin{document}

\ArticleTitle
  {Adapting coherent-state superpositions in noisy channels}
\ArticleAuthor*
  [0000-0002-5646-6964]
  {Jan Provazn\'{i}k}
  {provaznik@optics.upol.cz}
\ArticleAuthorAddress
  {Department of Optics, Palack\'{y} University, 17. listopadu 1192/12, 771 46 Olomouc, Czech Republic}

\ArticleAuthor
  [0000-0002-5761-8966]
  {Petr Marek}
\ArticleAuthorAddress
  {Department of Optics, Palack\'{y} University, 17. listopadu 1192/12, 771 46 Olomouc, Czech Republic}

\ArticleAuthor
  [0000-0001-8318-6514]
  {Julien Laurat}
\ArticleAuthorAddress
  {Laboratoire Kastler Brossel, Sorbonne Universit\'e, CNRS, ENS-Universit\'e PSL, Coll\`{e}ge de France, \\4 place Jussieu, 75005 Paris, France}

\ArticleAuthor
  [0000-0003-4114-6068]
  {Radim Filip}
\ArticleAuthorAddress
  {Department of Optics, Palack\'{y} University, 17. listopadu 1192/12, 771 46 Olomouc, Czech Republic}

\ArticleTitlePrint

\begin{abstract}\noindent
  Quantum non-Gaussian states are crucial for the fundamental understanding of non-linear bosonic systems and simultaneously advanced applications in quantum technologies. In many bosonic experiments the important quantum non-Gaussian feature is the negativity of the Wigner function, a cornerstone for quantum computation with bosons. Unfortunately, the negativities present in complex quantum states are extremely vulnerable to the effects of decoherence, such as energy loss, noise and dephasing, caused by the coupling to the environment, which is an unavoidable part of any experimental implementation. 
  An efficient way to mitigate its effects is by adapting quantum states into more resilient forms. We propose an optimal protection of superpositions of coherent states against a sequence of asymmetric thermal lossy channels by suitable squeezing operations. 
\end{abstract}

%
%

\FloatBarrier
\section{Introduction}

Fault tolerant quantum computation is one of the main goals of quantum information science~\cite{arute2019,wang2019,zhong2020}. Traveling light fields offer unprecedented scalability~\cite{rudolph2017,bourassa2021,inoue2023,asavanant2019,larsen2019} and have been already used to demonstrate a path towards quantum advantage~\cite{wang2019}. There, the information is encoded into non-Gaussian superpositions of quantum states~\cite{gottesman2001,menicucci2014,baragiola2019,pantaleoni2020}. Coherent {Schr\"{o}dinger} (CS) states, defined as quantum superposition of coherent states with different amplitudes can be considered an elementary version of such codes~\cite{ralph2003,lund2008,hastrup2022,rudolph2017,omkar2020}, and they can be also employed in quantum communication protocols~\cite{cerf2007,minzioni2019,guccione2020,darras2022}, quantum sensing~\cite{shinjo2021} and spectroscopy~\cite{milne2021}.

The main challenge faced by the optical systems is the decoherence, mainly due to optical loss and thermal noise. While high quality quantum states may offer the possibility of error correction~\cite{gottesman2001,menicucci2014,baragiola2019,pantaleoni2020}, decoherence poses severe limits on the preparation and propagation of such states~\cite{yukawa2013,eaton2022}. While energy loss can never completely remove certain quantum non-Gaussian features from quantum states~\cite{straka2014}, losing half of the signal to the environment is enough to completely obliterate any negativity of the Wigner function, thus preventing any possibility of quantum advantage~\cite{mari2012}. Loss combined with noise present in the environment makes its decay even faster.

It has been established that the effects of loss in photonic qubit systems can be probabilistically alleviated by suitable pre-processing of optical signals~\cite{micuda2012,gagatsos2014,chrzanowski2014} or channel engineering~\cite{starek2020}. In particular, CS states can be made more resilient to loss by applying a deterministic squeezing operation~\cite{filip2001,serafini2004,teh2020,filip2013,lejeannic2018,brewster2018,pan2023}. Specifically, it has been shown for pure lossy channels in~\cite{lejeannic2018}. However, to what extent can these methods be used for mitigation of added channel noise together with loss has not been studied so far.    

To address this question we consider a CS state propagating through a general series of channels incorporating loss and added phase-sensitive noise. We derive a general condition that needs to be satisfied to preserve any negativity present in the Wigner function of the quantum state based on optimal deterministic squeezing operations, that maximizes the central negativity of the Wigner function. In addition we extend the general condition to include even-parity CS states. This condition that must be satisfied for any negative values of the Wigner function to exist. 

Our analysis based primarily on negativity of the Wigner function is complemented by adapting the squeezing operations to maximize the Hilbert-Schmidt distance between CS states of opposite parities transmitted through the decohering channels.

%
%

\section{Minimizing decoherence in a single noisy channel}

%

Potential applications of CS states in quantum computation protocols range from their direct use~\cite{jeong2002,ralph2003,lund2008,hastrup2022,omkar2020,cerf2007}, through error correction~\cite{schlegel2022} to GKP state production, where they serve the important role of non-Gaussian building blocks~\cite{gottesman2001,weigand2018,vasconcelos2010}. 
Their versatility comes from their non-Gaussian features that diminish in the presence of Gaussian loss~\cite{filip2001,serafini2004,teh2020,filip2013,lejeannic2018,brewster2018,pan2023} and additive noise.

\begin{figure}[h]
  \begin{center}
    \includegraphics[width = 0.5 \columnwidth]{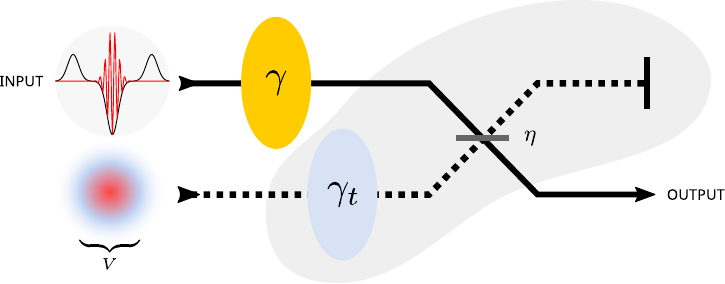}
  \end{center}
  \caption{  
    The signal (solid line) is transmitted through a lossy channel represented by a beam splitter with transmittance $\eta$ where it interacts with the environment (dashed line). The environment is assumed to be in an axis-aligned asymmetric thermal state, which can be interpreted as a symmetric thermal state characterized by its variance $V$ and the associated squeezing rate $\gamma_{t}$. The signal state can be protected against decoherence with an optional pre-squeezing operation. Its squeezing rate $\gamma$ can be adapted to offer the best protection of the transmitted CS state.
  }
  \label{fig-simple}
\end{figure}

Our search for methods capable of increasing their resilience against noise starts by considering a single lossy single channel with added noise, schematically depicted in \figref{fig-simple}. The channel can be modeled with an unbalanced beam splitter where the transmitted signal interacts with the environment in a generally asymmetrical thermal state, which can be understood as an initially symmetric thermal state that underwent squeezing. In our model we assume that the parameters of the channel are known.

The CS states propagating through the channel are defined as quantum superpositions of two coherent states with identical amplitudes and opposite phases. We recognize two types of CS states based on their parity,
\begin{equation}\label{css-ket}
  \ket{\xi, \pm} =
  \frac{1}{\sqrt{2}}
  \frac{
    \ket{\xi} \pm \ket{- \xi}
  }{
    \sqrt{1 \pm \exp( - 2 \abs{\xi}^2 )}
  }
\end{equation}
where $\ket{\xi}$ denotes coherent states --- eigenstates of the annihilation operator $\hat{a}$. The symbol $\ket{\xi, +}$ stands for the even-parity and $\ket{\xi, -}$ the odd-parity CS state. Their respective Wigner functions, defined in variables corresponding to quadrature operators with ${[\hat{x},\hat{p}] = \imath}$, read
\begin{equation}\label{css}
  \begin{aligned}
    W_{\xi,\pm}(X, P)
    & =
      \frac{1}{2}
      \frac{
        w_{\xi, \xi} (X, P)
        \pm \left[
          w_{\xi, -\xi} (X, P) +
          w_{-\xi, \xi} (X, P)
        \right] +
        w_{-\xi, -\xi} (X, P)
      }{1 \pm \exp( - 2 \abs{\xi}^{2} )} \Qd
  \end{aligned}
\end{equation}
They are expressed in terms of the complex-exponential building blocks
\begin{equation}\label{css-ws}
\begin{aligned}
  w_{\xi, \xi} (X, P) & = \frac{1}{\pi} \exp \left(
    - (X - X_{0})^{2}
    - (P - P_{0})^{2}
  \right) \Qc \\
  w_{\xi, -\xi} (X, P) & = \frac{1}{\pi} \exp \left(
    - (X - \imath P_{0})^{2}
    - (P + \imath X_{0})^{2}
  \right)
  \exp\left( - 2 \abs{\xi}^{2} \right) \Qc
\end{aligned}
\end{equation}
where we set ${\sqrt{2}\xi \equiv X_{0} + \imath P_{0}}$ and therefore ${2 \abs{\xi}^{2} \equiv X_{0}^{2} + P_{0}^{2}}$. As for the final building block, omitted in~\eqref{css-ws}, we have ${w_{- \xi, \xi} (X, P) \equiv \conj \{ w_{\xi, -\xi} (X, P) \}}$. These two complex-exponential blocks add up to a real-valued cosine. 


The Wigner function $W'_{\xi,\pm}(X', P')$ of the transmitted state is given by the integral transform
\begin{equation}\label{transform}
  W'_{\xi,\pm} (X', P') = \iint  W_{\xi,\pm}(X, P) K(X, P, X', P') \intd{X} \intd{P}
\end{equation}
of its initial Wigner function $W_{\xi,\pm} (X, P)$ with the kernel function $K(X, P, X', P')$ representing the quantum channel. The Gaussian channels under our consideration can be described with
\begin{equation}\label{kernel}
  K (X, P, X', P') =
  \frac { \pi^{-1} }{ \sqrt{ \sigma_{X} \sigma_{P} } }
  \exp \left(
    - \frac{ (X' - f_{X} X)^{2} }{ \sigma_{X} }
    - \frac{ (P' - f_{P} P)^{2} }{ \sigma_{P} }
  \right)
\end{equation}
where ${f_{X}, f_{P}, \sigma_{X}, \sigma_{P} \in \reals}$ parametrize the channel. Their exact values depend on the particular channel. The kernel function~\eqref{kernel} accommodates a broader range of channels beyond loss, including those where the transmitted states are squeezed before and after their interaction with the environment. It is, however, limited to channels preserving the separability of the quadrature variables. In our analysis we only consider channels with thermal environment, where the thermal state is axis-aligned with the transmitted CS state. This can be guaranteed by a using a phase-shift operation before their transmission.

When a CS state propagates through the channel~\eqref{kernel}, the blocks~\eqref{css-ws} transform into
\begin{equation}\label{css-ws-tf}
  \begin{aligned}
    w'_{\xi, \xi} (X', P') & = \frac{1}{\pi} \frac{1}{\sqrt{V_{X} V_{P}}} \exp \left(
      - \frac{(X' - f_{X} X_{0})^{2}}{V_{X}}
      - \frac{(P' - f_{P} P_{0})^{2}}{V_{P}}
    \right) \Qc \\
    w'_{\xi, -\xi} (X', P') & = \frac{1}{\pi} \frac{1}{\sqrt{V_{X} V_{P}}} \exp \left(
      - \frac{(X' - \imath f_{X} P_{0})^{2}}{V_{X}}
      - \frac{(P' + \imath f_{P} X_{0})^{2}}{V_{P}}
    \right)
    \exp\left( - 2 \abs{\xi}^{2} \right) \Qc
  \end{aligned}
\end{equation}
where ${V_{X} = \sigma_{X} + f_{X}^{2}}$ and ${V_{P} = \sigma_{P} + f_{P}^{2}}$. Together these form the transformed Wigner function
\begin{equation}\label{css-tf}
  \begin{aligned}
    W'_{\xi,\pm}(X', P')
      & =
      \frac{1}{2}
      \frac{
        w'_{\xi, \xi} (X', P')
        \pm \left[
          w'_{\xi, -\xi} (X', P') +
          w'_{-\xi, \xi} (X', P')
        \right] +
        w'_{-\xi, -\xi} (X', P')
      }{1 \pm \exp(- 2 \abs{\xi}^{2} )}
  \end{aligned}
\end{equation}
with structure similar to the initial Wigner function~\eqref{css} since both the integral kernel~\eqref{transform} and the complex-exponential fragments~\eqref{css-ws} are Gaussian functions separable in terms of $X$ and $P$.

It has been established~\cite{filip2013,lejeannic2018,brewster2018} that the rapid decay of negative values in Wigner functions of CS states due to pure loss (vacuum environment) can be mitigated with adaptive pre-squeezing operation prior to their transmission through the lossy channel. 

Channels of this kind, depicted in \figref{fig-simple}, can be characterized with parameters
\begin{equation}\label{kernel-SL}
  \begin{aligned}
    f_{X} & = \sqrt{\eta} \emath[- \gamma], &
    \sigma_{X} & = 2 (1 - \eta) \emath[-2 \gamma_{t}] V \Qc \\
    f_{P} & = \sqrt{\eta} \emath[+ \gamma], &
    \sigma_{P} & = 2 (1 - \eta) \emath[+2 \gamma_{t}] V \Qd
  \end{aligned}
\end{equation}
where $\gamma$ gives the squeezing rate of the pre-squeezing operation applied to the signal state, $\eta$ defines the (intensity) transmittance of the channel. The signal is mixed with the environment, an asymmetric thermal state, which can be parametrized by its initial symmetric variance $V$ and squeezing rate $\gamma_{t}$ of the asymmetrizing operation. Asymmetric thermal states remain classical for squeezing rates ${\abs{\gamma_{t}} \leq \log\sqrt{V}}$ as neither quadrature variance, ${V_{\pm} = \exp(\pm 2 \gamma_{t}) V}$, becomes squeezed below the vacuum threshold. The environment noise becomes non-classical beyond this limit.

Negative Wigner functions exclude Gaussian behavior of quantum states~\cite{filip2011,walschaers2021}. One of the prominent features of odd-parity CS states is the central negativity of their Wigner function. While central negativity is only a sufficient condition for quantum non-Gaussianity~\cite{walschaers2021}, its reduction can be used as a qualitative indicator of observability of the transmitted CS state above measurement and statistical noise~\cite{lejeannic2018,pan2023}. It also offers an experimental advantage as its measurement does not require a full state tomography~\cite{leibfried1996,bertet2002,laiho2010,kirchmair2013}. In optics, its value can be obtained directly as an expectation value of the photon number parity operator~\cite{banaszek1999} or computed from its photon number distribution~\cite{provaznik2020}.

Central negativity (CN) of the Wigner function of the odd-parity CS state is readily obtained from the transformed Wigner function~\eqref{css-tf} as
\begin{equation}\label{neg-odd}
  W_{\xi, -}'(0, 0) = m_{\xi, -}
  \Bigg\{
    \exp\left(
      - \frac{f_{X}^{2} }{V_{X}} X_{0}^{2}
      - \frac{f_{P}^{2} }{V_{P}} P_{0}^{2}
    \right) -
    \exp\left(
      - \frac{\sigma_{X}}{V_{X}} P_{0}^{2}
      - \frac{\sigma_{P}}{V_{P}} X_{0}^{2}
    \right)
  \Bigg\}
\end{equation}
where the normalization factor reads ${ m_{\xi, -}^{-1} = \pi \sqrt{V_{X} V_{P}} \left(1 - \exp(- 2\abs{\xi}^{2})  \right) }$. It will be negative iff
\begin{equation}\label{neg-odd-if}
  f_{X}^{2} f_{P}^{2} - \sigma_{X} \sigma_{P} > 0 \Qd
\end{equation}
This criterion can be straightforwardly generalized to include even-parity states. The transformed Wigner function~\eqref{css-tf} of the transmitted CS state, regardless of its parity, attains negative values if and only if the channel satisfies \eqref{neg-odd-if}. The necessary and sufficient condition 
\begin{equation}\label{neg-con}
  \exists (X, P) \in \reals^{2} : W'_{\xi, \pm} (X, P) < 0
  \iff
    f_{X}^2 f_{P}^2 - \sigma_{X} \sigma_{P} > 0 
\end{equation}
depends only on the parameters of the channel. If not satisfied, negative areas in the transmitted CS state, regardless of its parity or magnitude, can not be preserved by the discussed Gaussian squeezing operations. The conditon \eqref{neg-con} is proven within the supplementary material.

In the context of channels~\eqref{kernel-SL}, depicted in \figref{fig-simple}, the condition~\eqref{neg-con} provides a connection between the beam splitter transmittance $\eta$ and the symmetric variance $V$ of the thermal state. It establishes a region of feasible parameters of the lossy channel, independent of $\gamma_{t}$, bounded by
\begin{equation}\label{eta-V}
  \frac{2V}{1 + 2V} < \eta < 1
  \quad\text{and}\quad
  0 < V < \frac{\eta}{2 (1 - \eta)}
  \Qc
\end{equation}
where the negativities in the transmitted state survive. No amount of pre-squeezing $\gamma$ is going to protect CS states transmitted through channels~\eqref{kernel-SL} with parameters outside of this region.

\begin{figure}[h]
  \begin{center}
    \includegraphics[width = 0.9 \columnwidth]{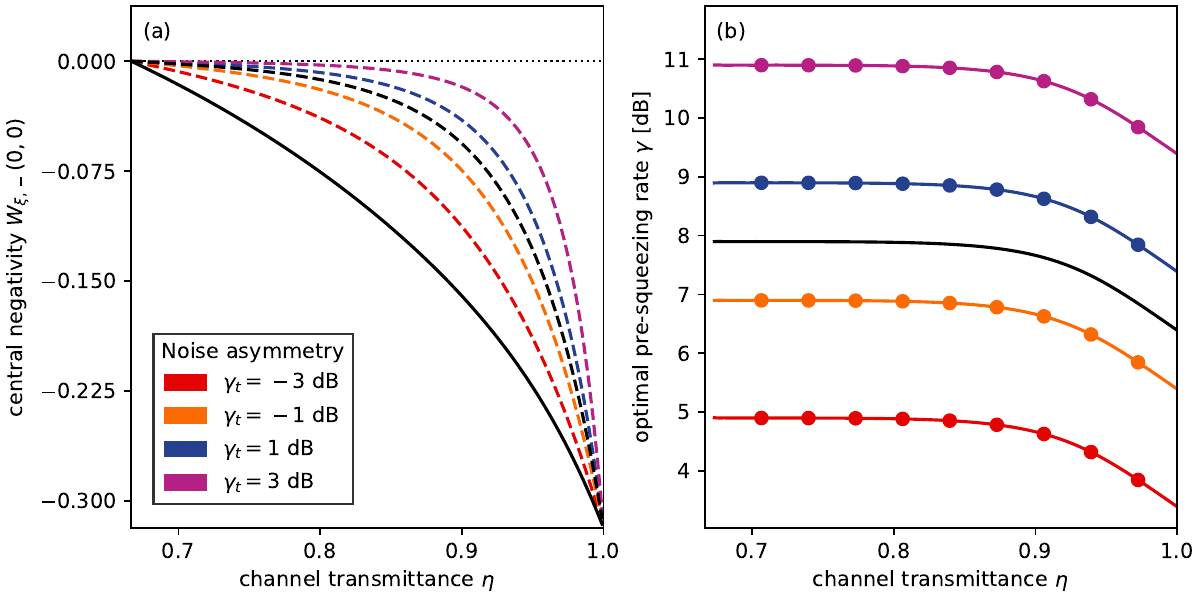}
  \end{center}
  \caption{
    Adaptive pre-squeezing protects odd-parity CS state (with ${\sqrt{2}\xi = 3}$) from decoherence due to loss and thermal noise with varying asymmetry.
    \textbf{(a)}~Dashed lines represent CN without protection. Asymmetries of the thermal noise with ${V = 1}$ (twice the variance of the vacuum state) are represented with different colors. We use black for the symmetric case with ${\gamma_{t} = 0}$, red for ${\gamma_{t} = 1 \dB}$, orange for ${\gamma_{t} = 3 \dB}$, blue for ${\gamma_{t} = 5 \dB}$, and purple for ${\gamma_{t} = 6 \dB}$ asymmetric cases.
    The solid black line represents the best attainable CN and does not depend on the $\gamma_{t}$ rate of the asymmetry.
    \textbf{(b)}~Optimal pre-squeezing rates $\gamma$ follow the same color scheme.  
    Colored lines represent optimal rates for different asymmetries. The lines appear constantly shifted by the value of $\gamma_{t}$ from the solid black line, which represents the optimal pre-squeezing $\gamma$ found for the symmetric thermal state (${\gamma_{t} = 0}$). The colored bullet points correspond to $\gamma_{t}$ added to its values at regular intervals to emphasise the constant shifts.
  }
  \label{fig-adapt-SL}
\end{figure}

In \figref{fig-adapt-SL} we consider lossy channels~\eqref{kernel-SL} with varying transmission rates and different asymmetries ${\gamma_{t} \in \{ -3, -1, 1, 3 \}\;\dB}$ of the channel noise characterized by ${V = 1}$, corresponding to a thermal state with double the variance of a vacuum state.
An odd-parity CS state with ${\sqrt{2} \xi = 3}$ is transmitted through the channel. The amplitude $\xi$ of the input CS state was chosen to be real number without a loss of generality as the state can be arbitrarily rotated in the phase space before its transmission.
The colored dashed lines in \figref{fig-adapt-SL}[a] represent CN of the state transmitted without any protection. The optimal squeezing rates of the protective pre-squeezing are determined by performing a numerical optimization of~\eqref{neg-odd} with respect to $\gamma$. The solid black line represents the best attainable CN when the state is optimally pre-squeezed prior to its transmission. It does not depend on the asymmetry $\gamma_{t}$ of the thermal state and coincides with the best attainable CN obtained for the symmetric thermal state.
The corresponding optimal squeezing rates $\gamma$ are shown in \figref{fig-adapt-SL}[b]. The solid black line represents the optimal pre-squeezing for the symmetric thermal state, whereas the colored solid lines represent different asymmetries of the state. The lines in the figure are shifted by constant offsets equal to the squeezing rate $\gamma_{t}$ of the asymmetry. This observation is supported by the colored bullet points which were obtained by adding the respective values of $\gamma_{t}$ to the optimal values of $\gamma$ obtained for the symmetric thermal state.

The optimal protective pre-squeezing operation can be interpreted as a pair of virtual squeezing operations with rates $\gamma_{0}$ and $\gamma_{1}$, joined together into a single physical squeezing operation with ${\gamma = \gamma_{0} + \gamma_{1}}$ rate. The first virtual squeezer, with $\gamma_{0}$, realizes the protective pre-squeezing of the transmitted CS state, while the other squeezer, with squeezing rate ${\gamma_{1} \equiv \gamma_{t}}$, reshapes the pre-squeezed state to match the asymmetry of the environment. Inverse squeezing operation can be applied after the channel to reshape the Wigner function back into its initial form without altering the non-Gaussian properties of the transmitted CS state. The optimal squeezing rate $\gamma_{0}$ of the protective pre-squeezing operation itself depends on the CS state, the excess of the thermal noise and the transmittance of the channel. It must be determined numerically on a case by case basis.

We can also interpret the reshaping of the state, performed by the second virtual squeezer, as a symmetrization of the environment. The second virtual squeezing operation $\gamma_{1}$ performed on the state and the virtual squeezing $\gamma_{t}$ of the thermal state share an identical squeezing rate. Consequently, the two squeezing operations can be propagated through the beam splitter in \figref{fig-simple}, thus reducing the resulting effective scheme to the actual protective pre-squeezing operation and a regular loss channel with a symmetric thermal noise. This interpretion allows us to focus solely on cases with symmetric noise.

\begin{figure}[h]
  \begin{center}
    \includegraphics[width = 0.9 \columnwidth]{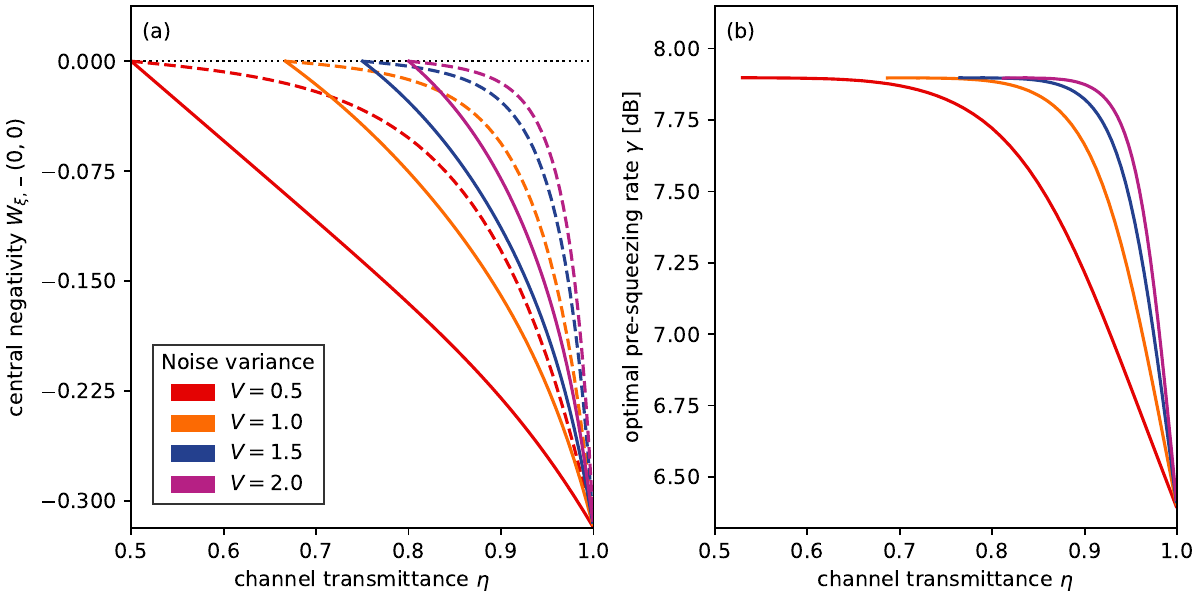}
  \end{center}
  \caption{
    Adaptive pre-squeezing protects odd-parity CS state (with ${\sqrt{2}\xi = 3}$) from decoherence due to loss and thermal noise. We consider symmetric thermal noise, characterized by its variance ${V \in \{ 0.5, 1.0, 1.5, 2.0 \}}$. 
    Colors are used to distinguish between individual variances.
    \textbf{(a)}~Dashed lines represent CN without protection. Solid lines correspond to CN attained by optimally pre-squeezed CS states.
    \textbf{(b)}~Optimal pre-squeezing rate $\gamma$ does not exhibit trivial dependence on the variance $V$ of the thermal noise.
  }
  \label{fig-adapt-SL-V}
\end{figure}

In \figref{fig-adapt-SL-V} we invesigate the adaptation of an odd-parity CS state with ${\sqrt{2}\xi = 3}$ transmitted through the lossy channel~\eqref{kernel-SL} with different amounts of symmetric thermal noise, characterized by its variance ${V \in \{ 0.5, 1.0, 1.5, 2.0 \}}$, where the thermal state with ${V = 0.5}$ corresponds to pure loss. In \figref{fig-adapt-SL-V}[a] we show the best attainable CN of odd-parity states. We employ the condition~\eqref{eta-V} to determine the least viable transmittance and only present the results where the transmitted CS state exhibits CN. The dashed lines represent cases without any protective pre-squeezing, wheres the solid solid lines, corresponding to the optimal adaptation of the transmitted CS state, exhibit significantly improved CN. In \figref{fig-adapt-SL-V}[b] we present the optimal squeezing rates $\gamma$ of the pre-squeezing operation. 
The optimal squeezing rate $\gamma$ does not exhibit trivial dependence on the variance $V$ and must be determined numerically on a case by case basis.

Adverse effects of asymmetric thermal noise can be mitigated with squeezing operations adapted to the asymmetry and the overall amount of the thermal noise in addition to adaptation for the loss rate and the transmitted CS state itself. The adaptation process can be understood as a combination of the protective pre-squeezing in the sense of~\cite{filip2013,lejeannic2018,brewster2018} and an additional squeezing operation, which effectively removes the asymmetry.
Consequently, in the analytical models used for the optimization, we can freely replace the lossy channels with asymmetrical thermal environment by lossy channels with equivalent symmetrical thermal noise.

Quantum states can propagate through diverse environments during their lifetime. To further our analysis we consider a natural extension to sequences of lossy quantum channels.

%
%

\FloatBarrier
\section{Minimizing decoherence across multiple noisy channels}

Composite quantum channels could prove to be advantageous by permitting complex mitigation strategies with protective operations potentially accessing the intermediate channel (channels) in the middle. Conversely, should no such complex mitigation protocol exist, adaptation of the state to the composite channel performed only at its beginning would still remain optimal.

A sequence of Gaussian channels~\eqref{kernel} can be concatenated together to form another Gaussian channel which can be described with the same formalism.  Similarly a Gaussian channel~\eqref{kernel} can be decomposed into more Gaussian channels~\cite{nicacio2010}. Simple decohering Gaussian channels have been extensively discussed in the literature~\cite{filip2013,lejeannic2018,brewster2018}, as well as in the previous section with some useful observations about symmetrization of the environment.

\begin{figure}[h]
  \begin{center}
    \includegraphics[width = 0.8 \columnwidth]{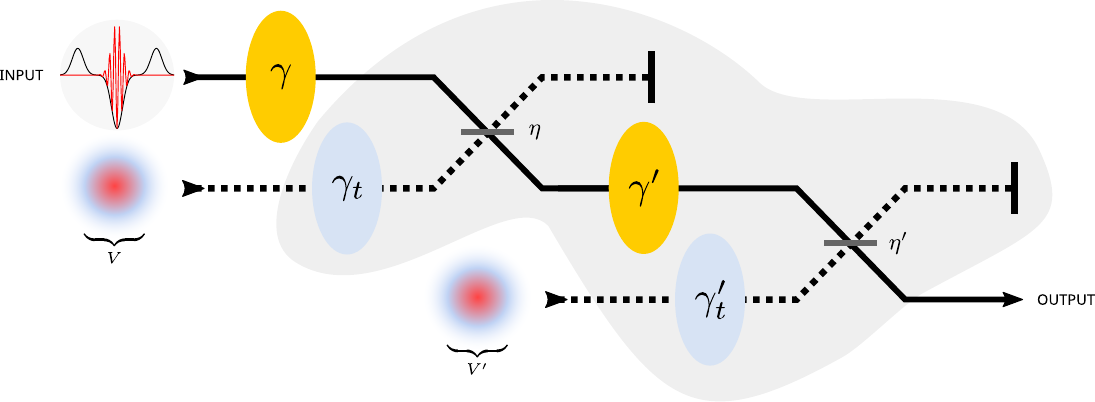}
  \end{center}
  \caption{
    Illustration of a composite channel comprising a pair of lossy channels with pre-squeezing ($\gamma$) and mid-squeezing ($\gamma'$). The first channel is parametrized by ($V, \gamma_{t}$) describing the asymmetric thermal state and $\eta$ determining its transmittance. Parameters of the second channel are distinguished by primes.
  }
  \label{fig-SLSL}
\end{figure}

Protection of CS states against decoherence may benefit from intermediate squeezing in composite channels. Consider the channel illustrated in \figref{fig-SLSL}. The signal CS state is pre-squeezed, passes through the first beam splitter. It is squeezed again before passing through the second beam splitter. It may be optionally squeezed as it leaves the channel, however, the final squeezing does not affect the central negativity of odd-parity CS states we use in our analysis. Its parametrization in terms of the formalism of~\eqref{kernel} is given by
\begin{equation}\label{kernel-SLSL}
  \begin{aligned}
      f_{X}
    & = \sqrt{\eta \eta'} \emath[- (\gamma + \gamma')] \Qc \\
      \sigma_{X}
    & = 2 (1 - \eta') \emath[- 2 \gamma'_{t}] V' +
        2 \eta' (1 - \eta) \emath[- 2 (\gamma' + \gamma_{t})] V \Qc \\
      f_{P}
    & = \sqrt{\eta \eta'} \emath[+ (\gamma + \gamma')] \Qc \\
      \sigma_{P}
    & = 2 (1 - \eta') \emath[+ 2 \gamma'_{t}] V' +
        2 \eta' (1 - \eta) \emath[+ 2 (\gamma' + \gamma_{t})] V' \Qc \\
  \end{aligned}
\end{equation}
where the transmittances of the channels are given by $\eta$ and $\eta'$, the pre-squeezing rate is $\gamma$ and the intermediate mid-squeezing rate is $\gamma'$.  The first thermal noise is described by the symmetric variance $V$ and the asymmetrizing squeezing $\gamma_{t}$. Similarly the second thermal noise is characterized by $V'$ and $\gamma'_{t}$.

\begin{figure}[h]
  \begin{center}
    \includegraphics[width = 1.0 \columnwidth]{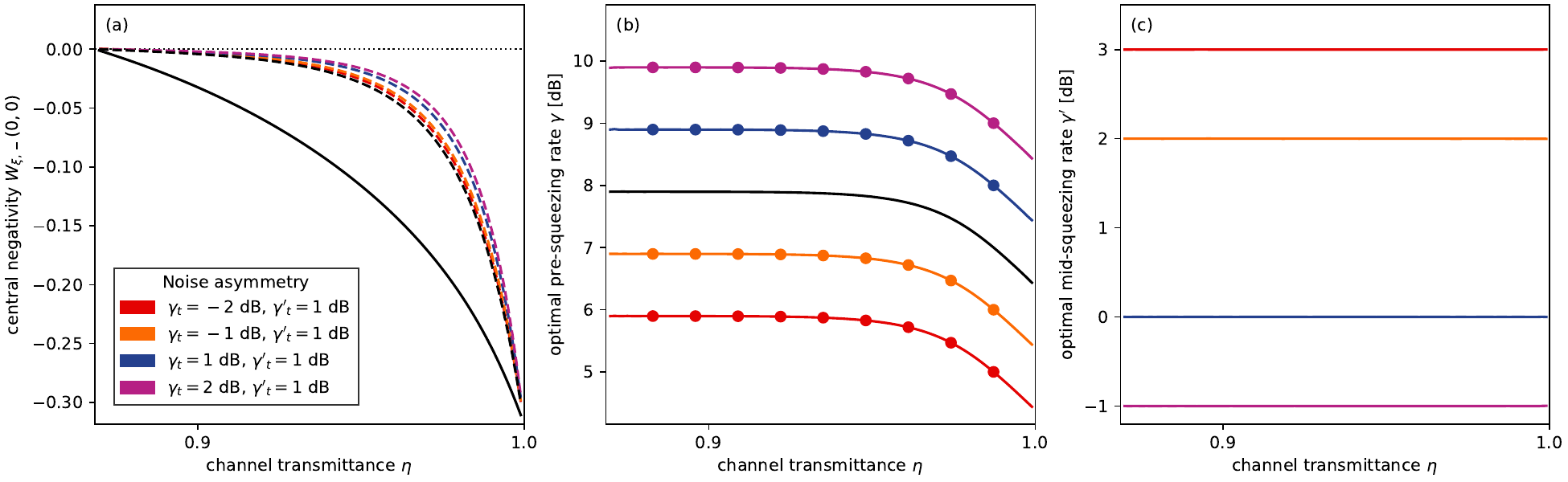}
  \end{center}
  \caption{
    Adaptive squeezing protects odd-parity CS state (${\sqrt{2} \xi = 3}$) from decoherence due to interaction with asymmetric thermal states with $V = 1$, ${\gamma_{t} = \{ -2, -1, 1, 2  \}\;\dB}$ in the first channel and ${V' = 2}$, ${\gamma_{t}' = 1 \dB}$ in the second channel.
    Solid black lines represent the cases where both thermal states are symmetric.
    \textbf{(a)}~Attainable central negativity. Dashed lines represent the attainable negativity without any adaptation. After adaptation, a part of which effectively symmetrizes the environment, the best attainable negativities coincide. The solid black line determines this best attainable negativity.
    \textbf{(b)}~Optimal pre-squeezing rate~$\gamma_{t}$ where the solid black line represents the optimal adaptation in the fundamental case when both thermals states of the environment are symmetric. Colored lines correspond to optimal pre-squeezing rates for asymmetric environments. These lines are shifted by a constant offset equal to $\gamma_{t}$. This fact is emphasized by the colored bullets that are obtained by adding~$\gamma_{t}$ to the fundamental pre-squeezing rate represented by the solid black line.
    \textbf{(c)}~Optimal mid-squeezing rate~$\gamma'_{t}$ depends only on the asymmetry of the adjacent thermal states. Its value is determined by the difference $\gamma'_{t} - \gamma_{t}$.
  }
  \label{fig-adapt-SLSL}
\end{figure}

In \figref{fig-adapt-SLSL} we explore the central negativity of the odd-parity CS state with ${\sqrt{2}\xi = 3}$ transmitted through a composite channel~\eqref{kernel-SLSL}. Without loss of generality we set the transmittances in both channels identical to keep the visualisation two dimensional. The plots show the best attainable negativity and the optimal pre-squeezing and mid-squeezing rates. We consider a number of asymmetric thermal environments. 
In \figref{fig-adapt-SLSL}[a] the dashed lines represent the attainable CN without any protective squeezing operations applied. The solid black line represents the best attainable CN when optimal pre-squeezing and mid-squeezing operations are applied to the CS state. The optimal CN depends only on the symmetric variance of the noise, rather than its asymmetry. The solid black line can be equally obtained for the same channel where both environments are symmetric. The solid black line in \figref{fig-adapt-SLSL}[b] represents the optimal pre-squeezing rate $\gamma$ obtained when there is no asymmetry in either thermal environment, whereas the colored lines correspond to different asymmetries. We recognize that these rates are constantly shifted from the symmetric case, the offset equal to $\gamma_{t}$ of the first channel. The colored bullets, obtained by adding the respective $\gamma_{t}$ rates to the black line, are used to highlight this observation. Furthermore, in the third plot, \figref{fig-adapt-SLSL}[c], we observe that the optimal mid-squeezing rates $\gamma'$ match the differences ${\gamma' \equiv \gamma'_{t} - \gamma_{t}}$ between the asymmetries of both environments.

We can interpret the pre-squeezing operation as a pair of virtual squeezing operations, as we did previously in the discussion of single channels, where the first virtual squeezer, with squeezing rate $\gamma_{0}$ realizes the actual protective pre-squeezing, whereas the second one reshapes the transmitted state to match the asymmetry of the environment in the first channel. The mid-squeezing operation can be also understood as a composition of two squeezers, where the first squeezer transforms the transmitted state back to its original shape, while the second one reshapes it once again to match the asymmetry of the environment within the second channel. Interestingly, we can also obtain the optimal mid-squeezing rate $\gamma'$ by maximizing the necessary condition~\eqref{neg-con}.

Because the optimal mid-squeezing rate depends solely on the asymmetries of individual environments, the composite channel~\eqref{kernel-SLSL} can be reduced into a single elementary channel~\eqref{kernel-SL} with effective transmittance $\eta_{e}$ and thermal noise variance $V_{e}$ given by
\begin{equation}\label{eff}
  \begin{aligned}
      \eta_{e}
    & = \eta \eta' \Qc \\
        V_{e}
    & = \frac{
      (1 - \eta') V' + (1 - \eta) \eta' V
    }{
      1 - \eta \eta'
    }
    \Qd
  \end{aligned}
\end{equation}
This brings the entire analysis of composite channels back to the previous discussion of elementary channels. The protection of the CS state is facilitated \textbf{entirely} by the initial pre-squeezing operation and composite channels~\eqref{kernel-SLSL} can be substituted with equivalent elementary channels~\eqref{kernel-SL} parametrized by~\eqref{eff}. Consequently, it is sufficient to only analyse noisy lossy channels with symmetric thermal environment.

With these conclusions in mind, we turn our attention back to the case of a single lossy channel with symmetric thermal noise and consider an alternative way of measuring the quality of the protection offered by the pre-squeezing operation.

\FloatBarrier
\section{Hilbert-Schmidt distance as a measure of state adaptation}

The appeal of the central negativity of the odd-parity CS state stems from its availability as it can be measured directly or straightforwardly estimated. The presence of negativity within the Wigner function is also a necessary condition for the presence of quantum non-Gaussianity in the transmitted state~\cite{filip2011,walschaers2021}, contextuality~\cite{booth2022} and advanced quantum protocols~\cite{chabaud2023}. It also establishes a bound on the rest of the negative regions of Wigner function of the odd-parity CS state. However, because it is a local measure, it does not provide sufficient information about the other negative regions of the Wigner function of the transmitted CS state and its other quantum non-Gaussian aspects.

To complete the analysis, we use a directly measurable~\cite{buhrman2001,filip2002,ekert2002,pregnell2006} Hilbert-Schmidt distance~\cite{ozawa2000} between opposite-parity CS states transmitted through the channel~\eqref{kernel-SL}. These two states form an orthonormal computational basis in quantum computation protocols~\cite{jeong2002,ralph2003,lund2008,hastrup2022,omkar2020,cerf2007}. Their orthogonality can be measured with the Hilbert-Schmidt distance; it deteriorates as their quantum non-Gaussian features decay due to decoherence.
While the distance itself does not directly indicate presence of quantum non-Gaussian features, our analysis determines whether the distance between the states can be improved through CS state adaptation. 

The Hilbert-Schmidt distance
\begin{equation}\label{alt-hs}
  \begin{aligned}
    \Delta_{\xi}
    & = 2 \pi \iint\limits_{\reals^{2}}
        \left[
          W'_{\xi, +} (X, P) - W'_{\xi, -} (X, P)
        \right]^{2}
        \intd{X} \intd{P} \\
    & = P_{\xi, +} - 2 Q_{\xi} + P_{\xi, -} 
  \end{aligned}
\end{equation}
is defined in terms of Wigner functions~\eqref{css-tf} and effectively pits the purities $P_{\xi, \pm}$ of the transmitted CS states against their mutual overlap $Q_{\xi}$. The distance is bounded, $0 \leq \Delta_{\xi} \leq 2$, both from below and above.
It can be expressed analytically. The final analytical expression is provided in the supplementary material.

\begin{figure}[h]
  \begin{center}
    \includegraphics[width = 0.9 \columnwidth]{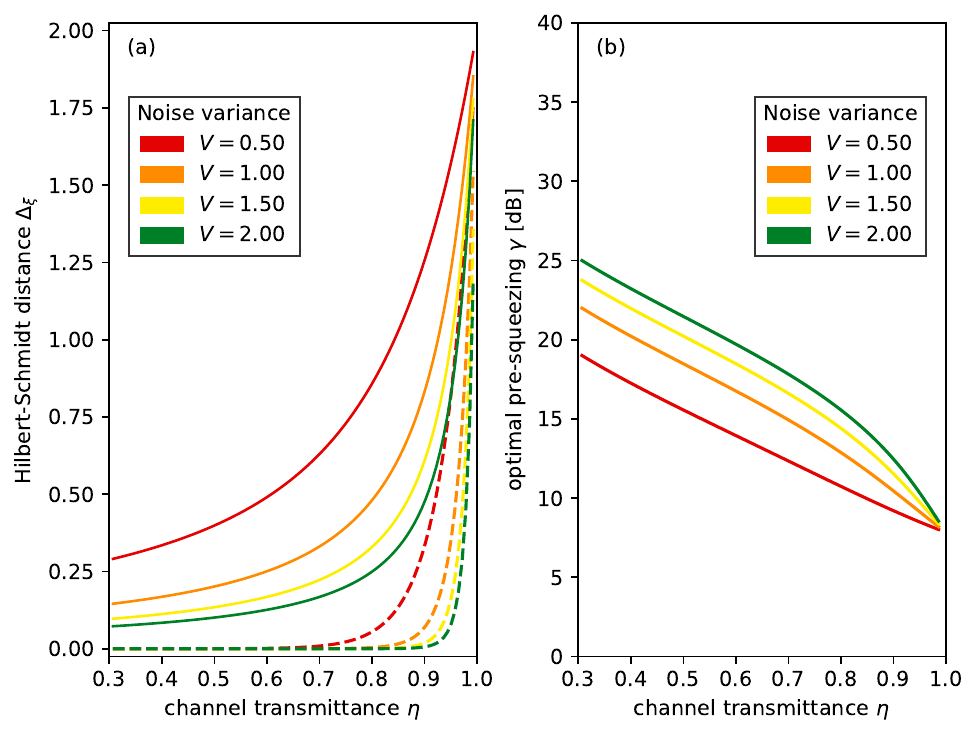}
  \end{center}
  \caption{
    Hilbert-Schmidt distance $\Delta_{\xi}$ between opposite-parity CS states (with ${\sqrt{2}\xi = 3}$) transmitted through the channel~\eqref{kernel-SL} where the environment is a symmetric thermal noise with variance ${V \in \{ 0.5, 1.0, 1.5, 2.0 \}}$. Colors indicate different thermal variance.
    \textbf{(a)}~The attainable Hilbert-Schmidt distance without protective pre-squeezing is represented with dashed lines, whereas the solid lines correspond to optimally pre-squeezed CS states. We can surmise that pre-squeezing certainly offers advantage as the attainable distance between the transmitted states is greater.
    \textbf{(b)}~The optimal pre-squeezing rate used to protect the CS states before their transmission.
  }
  \label{fig-alt-hs-x3p0-v2u}
\end{figure}

In \figref{fig-alt-hs-x3p0-v2u} we investigate the Hilbert-Schmidt distance for a particular CS state with ${\sqrt{2}\xi = 3}$ magnitude transmitted through an elementary channel~\eqref{kernel-SL}. It was established in prior sections that it is optimal to adapt the state to the asymmetry of the environment, or to equally symmetrize the environment. Consequently we consider only symmetric environment with distinct amount of thermal noise characterized by its variance ${V \in \{ 0.5, 1.0, 1.5, 2.0 \}}$ where ${V = 0.5}$ corresponds to pure loss.
The attainable Hilbert-Schmidt distance is presented in \figref{fig-alt-hs-x3p0-v2u}[a] where different colors encode different variances of the thermal noise. Dashed lines represent the distance obtained without the protective pre-squeezing operation, whereas the solid lines correspond to optimally pre-squeezed CS states. The adaptation is certainly advantageous as the distance between the transmitted CS states is lower without the protective pre-squeezing. The optimal pre-squeezing rates, presented in \figref{fig-alt-hs-x3p0-v2u}[b], do not depend trivially on the amount of the noise and must be determined numerically on a case by case basis.

The Hilbert-Schmidt distance between transmitted CS states with opposite parities can be used to measure the protective effects of the adaptive pre-squeezing operation. This measure successfully complements analysis based solely on central negativity as it takes both the even and odd parity CS states simultaneously into account.

We can see that suitable pre-squeezing operation can increase the distance of the states even in the cases when it is practically zero before the corrections. However, unlike the case of central negativity, where the optimal pre-squeezing did depend on the channel transmission only for its high values, here  the optimal pre-squeezing changes more or less linearly with the channel transmittance. This shows that there is no single universal guide to protecting the states as different aspects of the states require different methods of protection.

\FloatBarrier
\section{Conclusions and outlooks}

The loss and noise of bosonic channels deteriorating the non-Gaussian properties of quantum states is one of the main bottlenecks for scalable quantum computation with optical fields. While the noise can be, potentially, compensated by error correction~\cite{gottesman2001,lund2008,menicucci2014,baragiola2019,pantaleoni2020,hastrup2022}, this is costly and should come only after other, more feasible avenues, are explored first. We have presented such general mitigation strategy for superposed coherent states based only on feasible Gaussian operations. This operation can be realized either actively, by directly performing the Gaussian operation~\cite{miyata2014}, or it can be incorporated into the state preparation stage of the protocols~\cite{lejeannic2018}.

It can be straightforwardly extended to mitigate decoherence of the continuous variable component in quantum systems with hybrid entanglement between discrete qubits and continuous-variable CS states~\cite{cavailles2018,huang2019,lejeannic2018b,guccione2020}.

While the method is most valuable for optical fields propagating through bosonic channels, the concept is general and can be straightforwardly employed to preserve the non-Gaussian property of quantum states in other bosonic systems, such as microwave cavities, spin systems, trapped ions, or optomechanics~\cite{vlastakis2013,omran2019,chen2021,shomroni2020,tan2013,sun2021,fluhmann2019,miwa2014}.

%

\FloatBarrier
\section*{Acknowledgements}

The authors acknowledge early discussions with H. Le Jeannic. 
J.P. acknowledges fruitful discussions with O. Solodovnikova and J. Fiur\'{a}\v{s}ek.
J.P. acknowledges using the computational cluster at the Department of Optics at Palacký University and using several open-source software libraries~\cite{harris2020,virtanen2020,dalcin2021,hunter2007,meurer2017} in the computation and subsequent evaluation of presented results.


\FloatBarrier
\section*{Funding}

We acknowledge Grant No. 22-08772S of the Czech Science Foundation, the European Union's HORIZON Research and Innovation Actions under Grant Agreement no. 101080173 (CLUSTEC). 
This work was also supported by the French National Research Agency via the ShoQC Quantera project (ANR 19-QUAN-0005-05), and via the France 2030 projects QCommTestbed (ANR-22-PETQ-0011) and OQuLus (ANR-22-PETQ-0013). J.L. is a member of the Institut Universitaire de France.
In addition, J.P. acknowledges project IGA-PrF-2024-008 of Palacký University.

\FloatBarrier
\section*{References}
\printbibliography[heading = none]

%
%

\newpage
\supplement
\ArticleTitle
  {Supplementary material for\\Adapting coherent-state superpositions in noisy channels}
\ArticleTitlePrint

%
%

\section{Conditions for presence of negativity in Wigner functions}

It convenient to rewrite the complex-exponential form of the output Wigner function of the CS state into its manifestly real form,
\begin{equation}\label{sup-css-cos}
  \begin{aligned}
    W_{\xi, \pm}'( X', P') = \; & 2 m_{\xi, \pm}
    \exp\left(
      - \frac{X'^{2}}{V_{X}}
      - \frac{P'^{2}}{V_{P}}
    \right) 
    \Bigg\{ \\
      & \quad
      \exp\left(
        - \frac{f_{X}^{2} }{V_{X}} X_{0}^{2}
        - \frac{f_{P}^{2} }{V_{P}} P_{0}^{2}
      \right)
      \cosh \left( 2 \left[
        \frac{f_{X} }{V_{X}} X_{0} X' +
        \frac{f_{P} }{V_{P}} P_{0} P'
      \right] \right) \\
      & 
      \pm
      \exp\left(
        - \frac{\sigma_{X}}{V_{X}} P_{0}^{2}
        - \frac{\sigma_{P}}{V_{P}} X_{0}^{2}
      \right)
      \cos\left(2 \left[
        \frac{f_{P}}{V_{P}} X_{0} P' -
        \frac{f_{X}}{V_{X}} P_{0} X'
      \right]\right) \Bigg\} \Qc
  \end{aligned}
\end{equation}
where the factors $m_{\xi, \pm}$ stand for the normalization, given as
\begin{equation}
  m_{\xi, \pm}^{-1} = 
    2 \pi \sqrt{V_{X} V_{P}}
    \left(1 \pm \exp(- X_{0}^2 - P_{0}^2) \right)
  \Qd
\end{equation}
The coefficients $f_{X}, f_{P}, \sigma_{X}, \sigma_{P}, V_{X}$ and $V_{P}$ are all real and positive.

Let us begin with the \textbf{sufficient and necessary} condition for the existence of any negative area in the Wigner function of the odd-parity CS state. It is only natural to consider the central negativity in this case. The formula \eqref{sup-css-cos} simplifies into
\begin{equation}\label{sup-neg-odd}
  W_{\xi, -}'(0, 0) = 2 f_{-}
  \Bigg\{
    \exp\left(
      - \frac{f_{X}^{2} }{V_{X}} X_{0}^{2}
      - \frac{f_{P}^{2} }{V_{P}} P_{0}^{2}
    \right) -
    \exp\left(
      - \frac{\sigma_{X}}{V_{X}} P_{0}^{2}
      - \frac{\sigma_{P}}{V_{P}} X_{0}^{2}
    \right)
  \Bigg\}
\end{equation}
This expression will be negative if and only if 
\begin{equation}
  \begin{aligned}
    \exp\left(
      - \frac{f_{X}^{2} }{V_{X}} X_{0}^{2}
      - \frac{f_{P}^{2} }{V_{P}} P_{0}^{2}
    \right) -
    \exp\left(
      - \frac{\sigma_{X}}{V_{X}} P_{0}^{2}
      - \frac{\sigma_{P}}{V_{P}} X_{0}^{2}
    \right)
    < 0 \Qc
  \end{aligned}
\end{equation}
Direct comparison of the arguments within the exponential functions reveals the condition
\begin{equation}
  \begin{aligned}
    - X_{0}^{2} \left(
      \frac{ f_{X} f_{P} - \sigma_{X} \sigma_{P} }{ V_{X} V_{P} }
    \right)
    - P_{0}^{2} \left(
      \frac{ f_{X} f_{P} - \sigma_{X} \sigma_{P} }{ V_{X} V_{P} }
    \right) < 0
    \iff
    f_{X}^{2} f_{P}^{2} - \sigma_{X} \sigma_{P} > 0 
    \Qd
  \end{aligned}
\end{equation}
The central point of the odd-parity CS state will be negative if and only if ${ f_{X}^{2} f_{P}^{2} - \sigma_{X} \sigma_{P} > 0 }$. Notably this inequality does not depend on the state. It depends solely on the channel. This concludes the proof for the odd-parity state. \QED

The derivation of the condition for even-parity states is slightly more involved. We are looking for the \textbf{existence} of negative regions. This allows us to be somewhat liberal in the manipulation of the expression~\eqref{sup-css-cos}.

We begin with the observation that there are infinitely many $(X', P')$ pairs that satisfy
\begin{equation}\label{sup-neg-xp}
  \pm
  \cos\left(2 \left[
    \frac{f_{P}}{V_{P}} X_{0} P' -
    \frac{f_{X}}{V_{X}} P_{0} X'
  \right]\right) 
  \equiv 
  - \Abs{
      \cos\left(2 \left[
        \frac{f_{P}}{V_{P}} X_{0} P' -
        \frac{f_{X}}{V_{X}} P_{0} X'
      \right]\right) 
    }
  \Qc
\end{equation}
and, similarly, infinitely many $(X', P')$ pairs that also satisfy
\begin{equation}
  \cosh \left( 2 \left[
    \frac{f_{X} }{V_{X}} X_{0} X' +
    \frac{f_{P} }{V_{P}} P_{0} P'
  \right] \right) \equiv 1 \Qd
\end{equation}
The points ${(X', P')}$ satisfying both relations simultaneously can be determined analytically as 
\begin{equation}
  \begin{aligned}
    X'_{k} & = - \frac{V_{X}}{f_{X}} \frac{P_{0}}{X_{0}^{2} + P_{0}^{2}} \frac{2k + 1}{2} \pi \Qc \\
    P'_{k} & = + \frac{V_{P}}{f_{P}} \frac{X_{0}}{X_{0}^{2} + P_{0}^{2}} \frac{2k + 1}{2} \pi \Qc
  \end{aligned}
\end{equation}
by solving a pair of parametric linear equations with respect to $X'$ and $P'$,
\begin{equation}
  \begin{aligned}
    \frac{f_{X} }{V_{X}} X_{0} X' + \frac{f_{P} }{V_{P}} P_{0} P' & = 0 \Qc \\
    \frac{f_{P}}{V_{P}} X_{0} P' - \frac{f_{X}}{V_{X}} P_{0} X' & = \frac{2k + 1}{2} \pi \Qd
  \end{aligned}
\end{equation}
where the parameter ${\abs{k} \in \integers}$ embodies the periodicity of the cosine function. The Wigner function \eqref{sup-css-cos} simplifies into \eqref{sup-neg-odd} for these points and the proof reduces into the odd-parity case. \QED


%
%

\section{Closed form formula for the Hilbert-Schmidt distance}

We derive the closed form formula for the Hilbert-Schmidt distance~\cite{ozawa2000} between the opposite-parity CS states used in the paper. In particular, we have
\begin{equation}
  \begin{aligned}
    \Delta_{\xi}
    & = 2 \pi \iint\limits_{\reals^{2}}
        \left[
          W'_{\xi, +} (X', P') - W'_{\xi, -} (X', P')
        \right]^{2}
        \intd{X'} \intd{P'} \\
    & = P_{\xi, +} - 2 Q_{\xi} + P_{\xi, -} 
  \end{aligned}
\end{equation}
where the expressions for purity and the mutual overlap read
\begin{equation}
  \begin{aligned}
    P_{\xi, \pm}
    & = 2 \pi \iint\limits_{\reals^{2}} W_{\xi, \pm}^{2} (X, P) \intd{X} \intd{P} \\
    & = \frac{1}{ \sqrt{V_{X} V_{P}} }
        \frac{1 + M + N \pm 4 L}{
          2 \left(
            1 \pm \exp \left[ - X_{0}^{2} - P_{0}^{2} \right]
          \right)^{2}
        } \\
    Q_{\xi}
    & = 2 \pi \iint\limits_{\reals^{2}} W_{\xi, +} (X, P) W_{\xi, -} (X, P) \intd{X} \intd{P} \\
    & = \frac{1}{ \sqrt{V_{X} V_{P}} }
        \frac{1 + M - N}{
          2 \left(
            1 - \mu \exp \left[ - 2 X_{0}^{2} - 2 P_{0}^{2} \right]
          \right)
        } \Qd
  \end{aligned}
\end{equation}
The individual factors in both expressions were obtained directly by computing the underlying Gaussian integrals. Their closed forms are
\begin{equation}
  \begin{aligned}
    M & = \exp \left(
      - 2 \frac{ f_{X}^{2} }{ V_{X} } X_{0}^{2}
      - 2 \frac{ f_{P}^{2} }{ V_{P} } P_{0}^{2}
    \right) \\
    N & = \exp \left(
      - 2 X_{0}^{2} - 2 P_{0}^{2}
    \right) + \exp \left(
      - 2 \frac{ \sigma_{X} }{ V_{X} } P_{0}^{2}
      - 2 \frac{ \sigma_{P} }{ V_{P} } X_{0}^{2}
    \right) \\
    L & = \exp \left(
      - \left[
          \frac{ f_{X}^{2} }{ 2 V_{X} } +
          \frac{ f_{P}^{2} }{ 2 V_{P} } +
          \frac{ \sigma_{P} }{ V_{P} }
       \right] X_{0}^{2}
       - \left[
           \frac{ f_{X}^{2} }{ 2 V_{X} } +
           \frac{ f_{P}^{2} }{ 2 V_{P} } +
           \frac{ \sigma_{X} }{ V_{X} }
        \right] P_{0}^{2}
    \right) \times \\
    & \quad \cos \left(
      X_{0} \left[ \frac{ f_{X}^{2} }{ V_{X} } + \frac{ f_{P}^{2} }{ V_{P} } \right] P_{0}
    \right) 
  \end{aligned}
\end{equation}
where we use $V_{\bullet} = \sigma_{\bullet} + f_{\bullet}^{2}$ with $(\bullet \in \{ X, P \})$ for clarity and improved readability.

\FloatBarrier
\section*{References}
\printbibliography[heading = none]

\end{document}